\documentclass[12pt,a4paper]{iopart}
\usepackage{amssymb,graphicx,cite}
\eqnobysec
\begin{document}                    


 


 
 
   
 
\title[Jet formation in a collapsing Bose-Einstein condensate]{Mean-field
model of jet formation in a collapsing Bose-Einstein condensate}
 
\author{Sadhan K. Adhikari}
\address{Instituto de F\'{\i}sica Te\'orica, Universidade Estadual
Paulista, 01.405-900 S\~ao Paulo, S\~ao Paulo, Brazil\\}

\date{\today}
 
 
\begin{abstract}
 
We perform a systematic numerical study, based on the time-dependent
Gross-Pitaevskii equation, of jet formation in collapsing and exploding
Bose-Einstein condensates as in the experiment by Donley {\it et al.}
[2001 Nature {\bf 412} 295].  In the actual experiment, via a Feshbach
resonance, the scattering length of atomic interaction was suddenly
changed from positive to negative on a pre-formed condensate.
Consequently, the condensate collapsed and ejected atoms via explosion. On
a disruption of collapse by suddenly changing the scattering length to
zero a radial jet of atoms was formed in the experiment.  We present a
satisfactory account of jet formation under the experimental conditions as
well as make predictions beyond experimental conditions which can be
verified in future experiments.

\end{abstract}

\maketitle

\section{Introduction}
 
Recent successful observation   of Bose-Einstein condensates
(BEC) of trapped alkali atoms 
has initiated the intensive study of different novel phenomena.  
On the theoretical front, numerical simulation based on
the time-dependent nonlinear mean-field Gross-Pitaevskii (GP) equation
\cite{11} is well under control and 
has
provided a satisfactory account  of some  of these  phenomena.  

Since
the detection of BEC of $^7$Li atoms with attractive interaction, one
problem of extreme interest is the dynamical study of the formation and
decay of BEC for attractive atomic interaction \cite{ex2}. In general a
attractive condensate larger than a critical size is not dynamically
stable  \cite{ex2}. However, if such a 
condensate is ``prepared"  or somehow made to exist it experiences a
dramatic collapse and explodes 
emitting atoms.

A dynamical study of the collapse   
has been performed by Donley et al.
\cite{don} on an attractive  $^{85}$Rb BEC \cite{ex3}
in an axially
symmetric trap, where they
manipulated the inter-atomic interaction by changing the external magnetic
field exploiting  a nearby Feshbach resonance \cite{fs}. 
In the vicinity
of a Feshbach resonance the atomic scattering length $a$ can be varied
over a huge range by adjusting the external magnetic field.  
Consequently,
they  changed the sign of the scattering length, thus
transforming a
repulsive condensate of $^{85}$Rb atoms 
into an attractive one which naturally evolves into a collapsing and
exploding condensate. 
Donley et al.
have provided a quantitative estimate of the explosion of this BEC
by measuring different properties of the exploding condensate.

It has been realized that many features of the experiment by Donley {\it
et al.} \cite{don} can be described by the mean-field GP equation
\cite{th1,th2,th2a,th3,th3a,th4,th5,th6,th7,th8,th9,th10}. 
However, we are fully aware
that there are features of this experiment which are expected to be beyond
mean-field
description. Among these are the distribution of number  and energy  of
emitted
high-energy ($\sim 10^{-7}$  Kelvin)
uncondensed burst atoms reported in the experiment. Although there have
been some attempts \cite{th3,th3a,th4}
to describe the burst atoms using the mean-field GP
equation, now there  seems to be a consensus 
that they cannot be
described
adequately and satisfactorily 
using
a mean-field approach
\cite{th5,th6,th7,th8,th9,th10}. 
 The GP
equation is supposed to deal with the zero- or very 
low-energy condensed phase of atoms
and has been successfully used to predict the time to collapse, evolution
of the collapsing   condensate as well as its
oscillation \cite{th2,th2a,th3,th4,th5}. However, the GP equation has not
been fully
tested  
to study
the very low-energy ($\sim$ nano Kelvin) jet formation \cite{don}
when the collapse
is suddenly stopped before completion by jumping the scattering length to
zero (noninteracting atoms) or positive (repulsive atoms) values. As the
jet atoms are very low-energy condensed atoms  the mean-field GP
equation seems to be suitable for their study and we present
such a systematic description in this paper.

In the experiment \cite{don} the initial scattering length
$a_{\mbox{initial}}$ ($\ge 0$)  of a repulsive ($a>0$)
or noninteracting ($a=0$)  condensate
is suddenly
jumped to the  negative value $a_{\mbox{collapse}}$ ($< 0$) to start the
collapse. The condensate then begins to collapse and explode. The collapse
is then suddenly  terminated   after an interval of time
$t_{\mbox{evolve}}$
by
jumping the scattering length from  $a_{\mbox{collapse}}$ to
$a_{\mbox{quench}}$  ($\ge 0$).  
The jet atoms are slowly formed in the radial
direction
when
the collapse is stopped  in this fashion. In the experiment usually
$a_{\mbox{quench}}=0$. Sometimes the scattering length is jumped past
$a_{\mbox{quench}}$ to $a_{\mbox{expand}} = 250a_0$ to have an expanded
condensate to facilitate observation. It is emphasized that unlike the
emitted 
uncondensed ``hotter"
missing
and burst atoms reported in the experiment \cite{don} the jet atoms form a
part of the surviving ``colder" condensate and hence should be describable
by
the mean-field GP equation. Yet most of the theoretical treatments on the
topic \cite{th2,th2a,th4,th5,th6,th7,th8} are completely silent about jet
formation.  Saito {\it et al.}  \cite{th3} and Bao {\it et al.}
\cite{th10} present a mean-field description of jet formation and Calzetta
{\it et al.}       \cite{th9}  goes beyond the mean-field model in
including the effects of quantum field corrections  in their description
of jet formation. Bao {\it et al.}   \cite{th10} employ a fully
asymmetric mean-field model and are capable of studying the breakdown of
axial symmetry in jet formation for experiments of collapse performed 
in an axially symmetric trap \cite{don}.

Although, the breakdown of axial symmetry     and the possible
necessity of  quantum field corrections  in the jet formation are
interesting topics to be studied carefully, we investigate the possibility
of  explaining the jet formation within an axially-symmetric mean-field
model.
Using the GP equation
we study satisfactorily 
the jet formation  for the experimental values  of the
scattering lengths and times which shows that a mean-field model
describes the essential features of jet formation.
  The number of jet atoms is in agreement with
experiment. Further, we extend our study to other values of
scattering lengths and times and predict the possibility of the formation 
of jet atoms. Future experiments may test these predictions and thus
provide a more stringent test for the mean-field GP equation. 
To account for the loss of atoms from the
strongly attractive collapsing condensate we include an absorptive
nonlinear three-body recombination term in the GP equation
\cite{th1}.

In section 2 we present our  mean-field model. In section 3 we present
our results that we compare with the
experiment and other numerical studies. In section 4  we present a
physical
discussion of our findings 
and some concluding remarks  are given in section 5.
 
\section{Nonlinear Gross-Pitaevskii Equation}
 
 
The time-dependent Bose-Einstein condensate wave
function $\Psi({\bf r};\tau)$ at position ${\bf r}$ and time $\tau $
allowing
for atomic loss
may
be described by the following  mean-field nonlinear GP equation
\cite{11}
\begin{eqnarray}\label{a} \biggr[& -& i\hbar\frac{\partial
}{\partial \tau}
-\frac{\hbar^2\nabla^2   }{2m}
+ V({\bf r})
+ gN|\Psi({\bf
r};\tau)|^2-  \frac{i\hbar}{2}
\nonumber \\
& \times & (K_2N|\Psi({\bf r};\tau) |^2
+K_3N^2|\Psi({\bf r};\tau) |^4)
 \biggr]\Psi({\bf r};\tau)=0.
\end{eqnarray}
Here $m$
is
the mass and  $N$ the number of atoms in the
condensate,
 $g=4\pi \hbar^2 a/m $ the strength of inter-atomic interaction, with
$a$ the atomic scattering length. 
The terms $K_2$ and $K_3$ 
denote two-body
dipolar and three-body recombination loss-rate coefficients, respectively
and include the Bose statistical factors 
$1/2!$ and $1/3!$ needed to describe the condensate.
The trap potential with cylindrical symmetry may be written as  $  V({\bf
r}) =\frac{1}{2}m \omega ^2(r^2+\lambda^2 z^2)$ where
 $\omega$ is the angular frequency
in the radial direction $r$ and
$\lambda \omega$ that in  the
axial direction $z$. 
The normalization condition of the wave
function is
$ \int d{\bf r} |\Psi({\bf r};\tau)|^2 = 1. $
Here we
simulate the atom  loss via
the most important quintic three-body term  $K_3$ \cite{th1,th2,th2a,th3}.
The contribution of the cubic  two-body  loss term
\cite{k3} is
expected to be negligible \cite{th1,th3} compared to the  three-body term
in
the present problem of the  collapsed condensate with large density
and will not be considered here.

In the absence of angular
momentum the wave function has the form $\Psi({\bf
r};\tau)=\psi(r,z;\tau).$
Now  transforming to
dimensionless variables
defined by $x =\sqrt 2 r/l$,  $y=\sqrt 2 z/l$,   $t=\tau \omega, $
$l\equiv \sqrt {\hbar/(m\omega)}$,
and
\begin{equation}\label{wf}
\phi(x,y;t)\equiv
\frac{ \varphi(x,y;t)}{x} =  \sqrt{\frac{l^3}{\sqrt 8}}\psi(r,z;\tau),
\end{equation}
we get
\begin{eqnarray}\label{d1}
\biggr[-i\frac{\partial
}{\partial t} -\frac{\partial^2}{\partial
x^2}+\frac{1}{x}\frac{\partial}{\partial x} -\frac{\partial^2}{\partial
y^2}
+\frac{1}{4}\left(x^2+\lambda^2 y^2-\frac{4}{x^2}\right) \nonumber \\
+ 8 \sqrt 2 \pi   n\left|\frac {\varphi({x,y};t)}{x}\right|^2
- i\xi n^2\left|\frac {\varphi({x,y};t)}{x}\right|^4
 \biggr]\varphi({ x,y};t)=0,
\end{eqnarray}
where
$ n =   N a /l$
and $\xi=4K_3/(a^2l^4\omega).$
The normalization condition  of the wave
function becomes
\begin{equation}\label{5} {\cal N}_{\mbox{norm}}\equiv {2\pi} \int_0
^\infty
dx \int _{-\infty}^\infty dy|\varphi(x,y;t)|
^2 x^{-1}.  \end{equation}
For $K_3=0,$  ${\cal N}_{\mbox{norm}}=1$, however, in the presence of loss
$K_3 > 0$, ${\cal N}_{\mbox{norm}}
< 1.$ The number of remaining atoms $N$
in the condensate is given by $ N=N_0
{\cal N}_{\mbox{norm}}$, where $N_0$ is the initial number of atoms.

In this study the term $K_3$ will be used for a description of atom loss
in the case of attractive interaction.  Near a Feshbach resonance the
variation of $K_3$ vs. scattering length $a$ is very rapid and complicated
\cite{esry}. From theoretical \cite{ver} and experimental \cite{k3}
studies it has been found that for negative $a,$ $K_3$ increases rapidly
as $|a|^n$, where the theoretical study \cite{ver} favors $n=2$ for
smaller values of $|a|$. In this work we represent this variation via a
quadratic dependence: $K_3\sim a^2$. This makes the
only ``parameter"  $\xi$
of the present model
a constant for an experimental set up with fixed $l$ and $\omega$ and in
the present study we use a constant $\xi$. 
In our previous and the present investigation we
choose  $\xi$ or $K_3$ to provide a correct evolution of the number of
atoms in the  condensate during collapse and explosion. The mean-field GP
equation is best-suited to make this prediction.  After a small experimentation
it is found that $\xi=2$ fits the time evolution of the
condensate in the experiment of Donley {\it et al.} \cite{don} 
satisfactorily for a wide range of variation of initial number of atoms
and scattering lengths
\cite{th2}.
This value of $\xi$ is used in all   simulations  reported
in this paper. A similar philosophy is used in choosing the value of $K_3$
in \cite{th5,th10}, where the authors reproduced the rate of variation of
the number of atoms in the collapsing condensate. However, interest in 
other mean-field
\cite{th3,th4}
and beyond-mean-field \cite{th6,th7,th8}  
studies of the collapsing condensate was not in 
reproducing the variation of
the number of atoms in the collapsing condensate. Consequently, 
if a three-body recombination term is used in these studies other
criteria are used in fixing the value of $K_3$.

It is useful to compare this value of $\xi (=2)$ with the experimental
\cite{k3}
estimate of three-body loss rate of
$^{85}$Rb as well as with other values used in the study of the
experiment by Donley {\it et al.}  For this we
recall that the present value  $\xi =2$  with 
$K_3=\xi a^2l^4\omega/4$  leads to \cite{th2,th2a}
$K_3\simeq 8\times 10^{-25}$ cm$^6$/s at $a=-340a_0$ and  
$K_3\simeq 6\times 10^{-27}$ cm$^6$/s at $a=-30a_0$.  
Santos {\it et al.} \cite{th4}
employed the experimental value \cite{k3}
$K_3\simeq 7\times 10^{-25}$ cm$^6$/s at $a=-340a_0$ which is
very close to the present choice. (Santos {\it et
al.}   
quote $L_3 \equiv 6K_3$. We overlooked this fact in \cite{th2,th2a} and
hence
misquoted there the $K_3$ values of \cite{th4,k3}.)
Bao {\it et al.} \cite{th10}
employed 
$K_3\simeq 6.75\times 10^{-27}$ cm$^6$/s at $a=-30a_0$ in close 
agreement with the present choice, which leads to a very similar time 
evolution of the number of atoms in the collapsing condensate. 
Savage {\it et al.} \cite{th5}
employed a slightly larger value 
$K_3\simeq 19\times 10^{-27}$ cm$^6$/s at $a=-30a_0$ and also produced
reasonably
similar results for the  time
evolution of the number of atoms. However, 
Saito  {\it et al.} \cite{th3} and Duine  {\it et al.} \cite{th7} 
employed a much smaller value (smaller by more than an order of magnitude)
$K_3\simeq 2\times 10^{-28}$ cm$^6$/s at $a=-30a_0$. With this value of
$K_3$, unlike in the other studies \cite{th4,th5,th10,th2}, 
it is not possible to fit the  
the  time evolution of the number of atoms in the collapsing condensate.  
Of these theoretical studies, the $K_3$ values used by Santos {\it et al.}
\cite{th4},  Savage {\it et al.} \cite{th5},
Bao 
{\it et
al.} \cite{th10} and the present author  \cite{th2} are
consistent with each other 
and describes well the decay of the collapsing condensate.

\section{Numerical Result}

\begin{figure}[!ht]
 
\begin{center}
\includegraphics[width=0.48\linewidth]{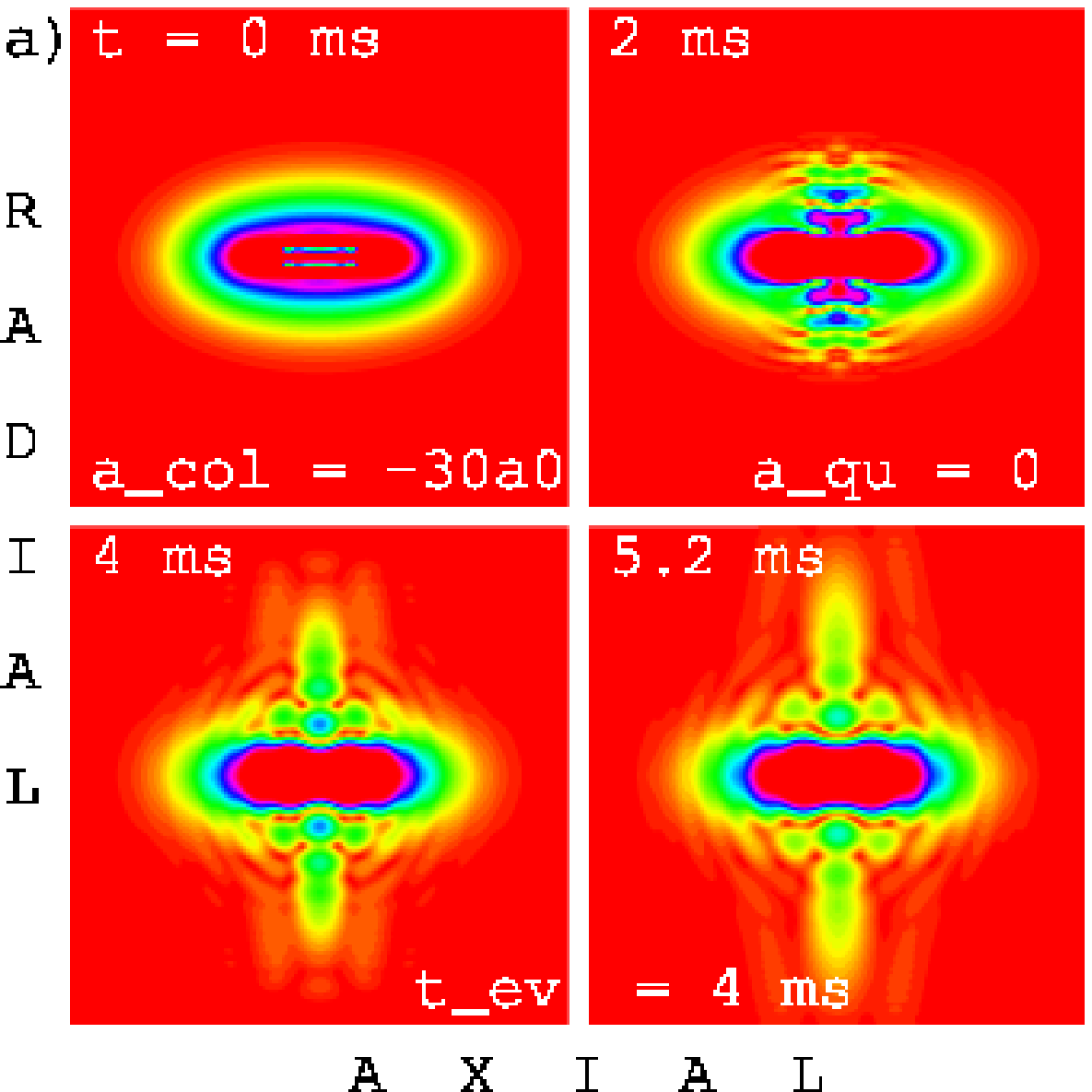}
\includegraphics[width=0.48\linewidth]{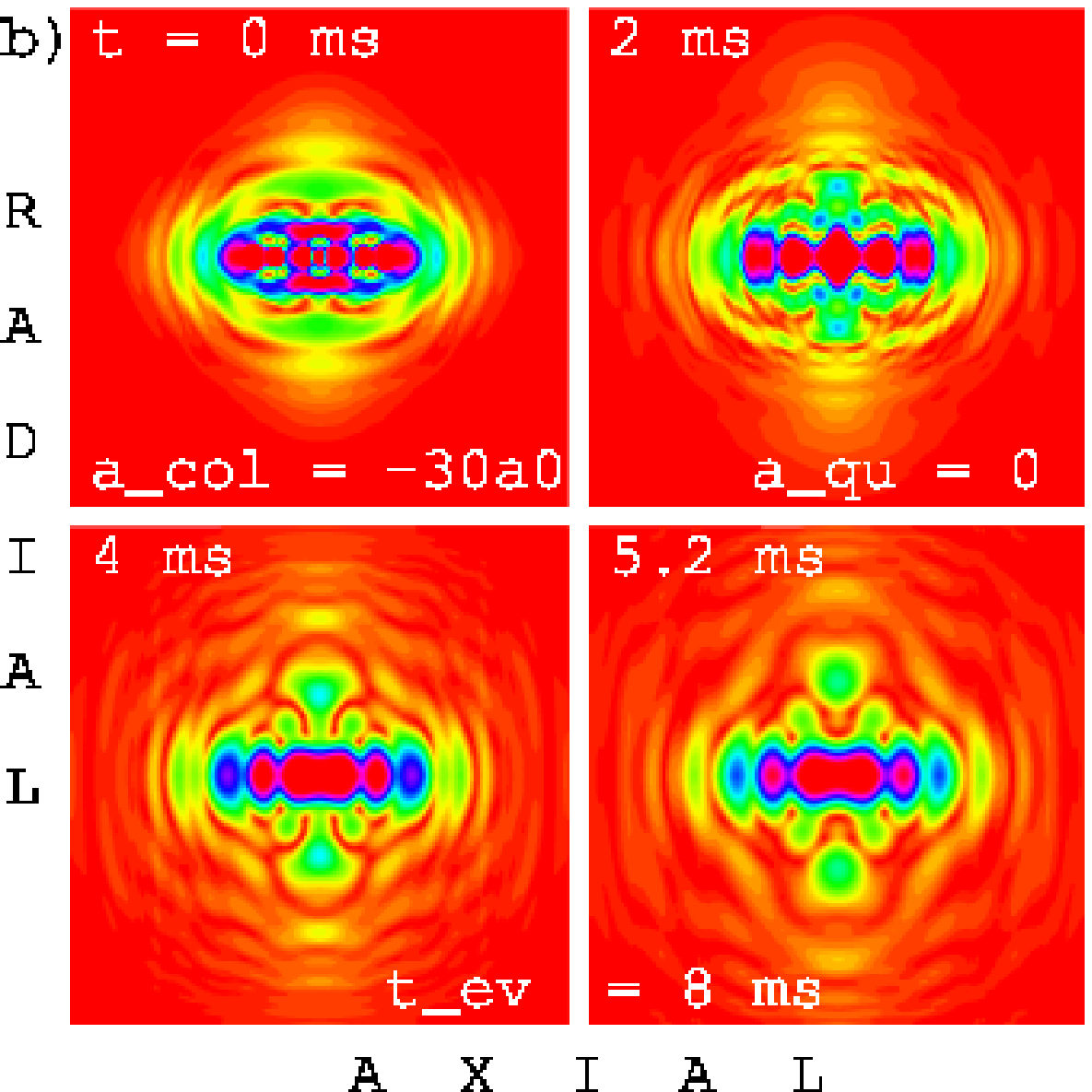}
\end{center}

\caption{A view of the evolution of radial jet at times $t=0$, 2ms, 4ms
and
5.2 ms on a mat of size 16 $\mu$m $\times$ 16 $\mu$m from a contour plot
of $|\phi(x,y,t)|^2$ 
for  
$a_{\mbox{initial}}=7a_0$, $a_{\mbox{collapse}}=-30a_0$, $\xi=2$, 
$N_0=16000$, $a_{\mbox{quench}}= 0$,  and (a) 
$t_{\mbox{evolve}} = 4$ ms and (b) $t_{\mbox{evolve}} = 8$ ms.}

\end{figure}
 
We solve the GP equation (\ref{d1}) numerically  using a time-iteration
method based on the Crank-Nicholson discretization scheme
elaborated in  \cite{sk1}.  
We
discretize the GP equation
using time step $\Delta=0.001$ and space step $0.05$ for both
$x$ and $y$ spanning $x$ from 0 to 15 and $y$ from $-40$ to 40. This
domain of space was sufficient to encompass  the whole condensate wave
function in this study.

The calculation is performed with the actual parameters of the experiment
by Donley {\it et al.} \cite{don}, e. g,, the initial number of atoms,
scattering lengths, etc. 
The numerical simulation using (\ref{d1}) with a nonzero $\xi (=2)$
immediately yields the remaining number of atoms in the condensate
after the jump in scattering length.
The remaining  number of atoms vs. time  for
$a_{\mbox{initial}}=7a_0$, $a_{\mbox{collapse}}=-30a_0$, $\xi=2$, and
$N_0=16000$ is in satisfactory agreement with experiment 
\cite{th2}. 

Now we
consider the jet formation as in the experiment for these sets of the
parameters after different
evolution times $t_{\mbox{evolve}}$ of the collapsing condensate when the
scattering length is suddenly changed from  $a_{\mbox{collapse}}=-30a_0$
to $a_{\mbox{quench}}= 0$ or to $250a_0$. First we consider
$a_{\mbox{quench}}= 0$. In figures 1 (a) and (b)  we plot the  contour
plot of the
condensate for $t_{\mbox{evolve}} = 4$ ms and 8 ms, respectively, 
in this case at
different times $t=0$, 2 ms, 4 ms, and 5.2 ms after jumping the
scattering length to $a_{\mbox{quench}}= 0$.  
A prominent radial jet is formed at time $t=5.2$ ms after stopping the
collapse for
$t_{\mbox{evolve}} = 4$ ms. The jet is formed slowly after  stopping the
collapse and is more prominent 4 $-$ 6 ms  after stopping the collapse.
The jet for  $t_{\mbox{evolve}} = 8$ ms is not
so spectacular. We also studied the jet formation for  $t_{\mbox{evolve}}
=
2$ ms, 6 ms, and 10 ms. The jet is much less pronounced for
$t_{\mbox{evolve}}
=
2$ ms  and 10 ms compared to the jet in figures 1.

Next we study the effect of the variation of $a_{\mbox{collapse}}=-30a_0$
on
the jet formation. For this purpose for the same set of parameters of
figures 1 
we consider
$a_{\mbox{collapse}}=-6.7a_0$  and  $a_{\mbox{collapse}}=-250a_0$
corresponding to smaller and larger attraction, in figures 2
(a) and (b), respectively. First, we consider the case
$a_{\mbox{collapse}}=-6.7a_0$ in figure 2 (a). In this case the final
attraction is weaker and the collapse is less dramatic.  The collapse and
the decay of
atoms do not start until  $t_{\mbox{evolve}}\approx 6$ ms. For
$t_{\mbox{evolve}}\approx 8$ ms, a broad jet is formed after about 12 ms 
of disruption of collapse. After 4 ms of disruption of collapse there is
almost no visible jet. The jets are broader and less prominent for larger 
values of $t_{\mbox{evolve}}$. Next we consider the highly attractive case 
$a_{\mbox{collapse}}=-250a_0$ in figure 2 (b). In this case due to a very
large
attraction the collapse and the decay of
atoms start at a small value of  $t_{\mbox{evolve}}$ close to 
zero. Hence a reasonable jet is formed for  $t_{\mbox{evolve}}=2$
ms at small times after stopping the collapse. Because of large attraction
in this case the collapse
is over for a
smaller value of   $t_{\mbox{evolve}}$. Hence there is almost no jet
formation for
$t_{\mbox{evolve}}\ge 4$ ms. However, the nature of jets in each
case of figures 1 and
2 is distinct.

\begin{figure}
 
\begin{center}
\includegraphics[width=0.48\linewidth]{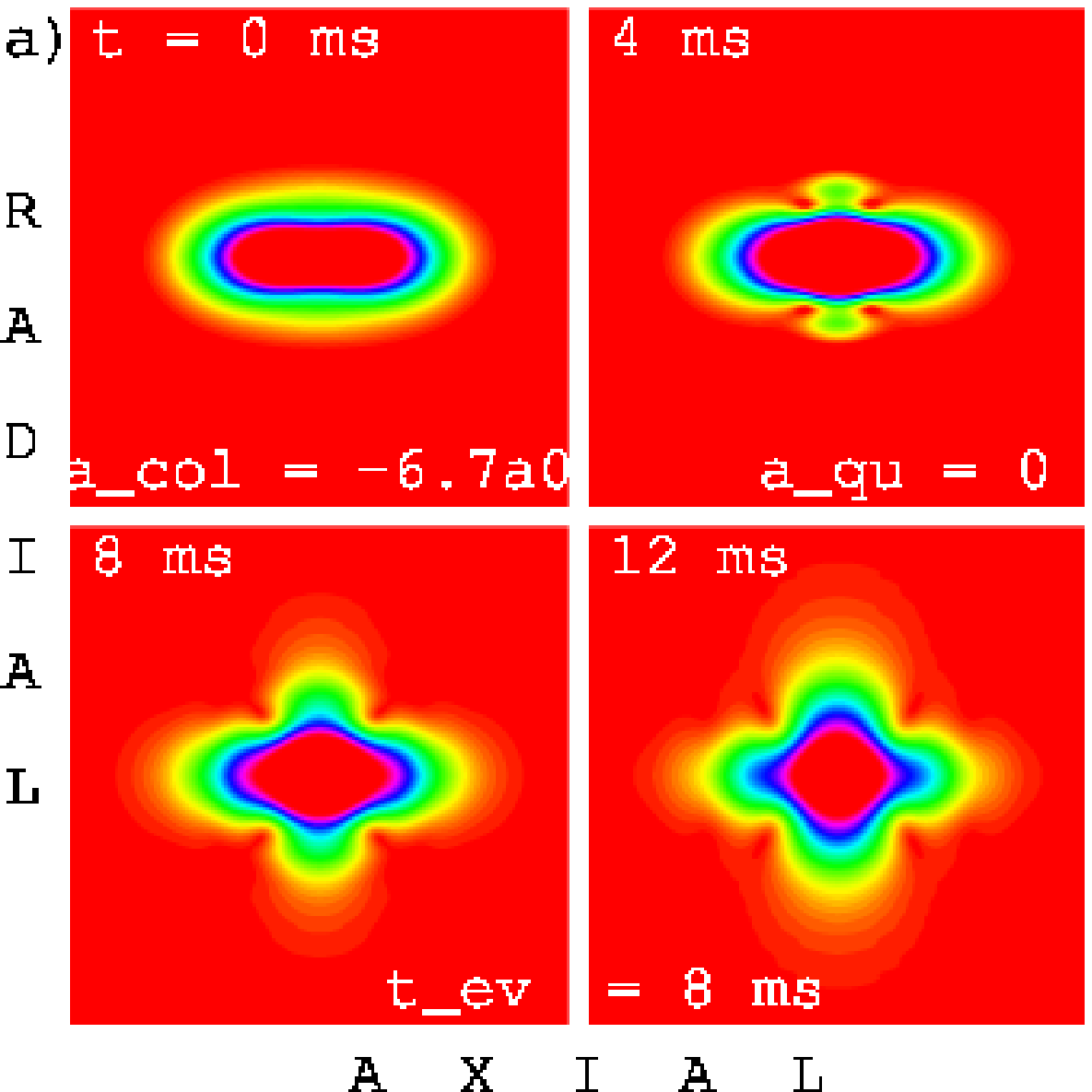}
\includegraphics[width=0.48\linewidth]{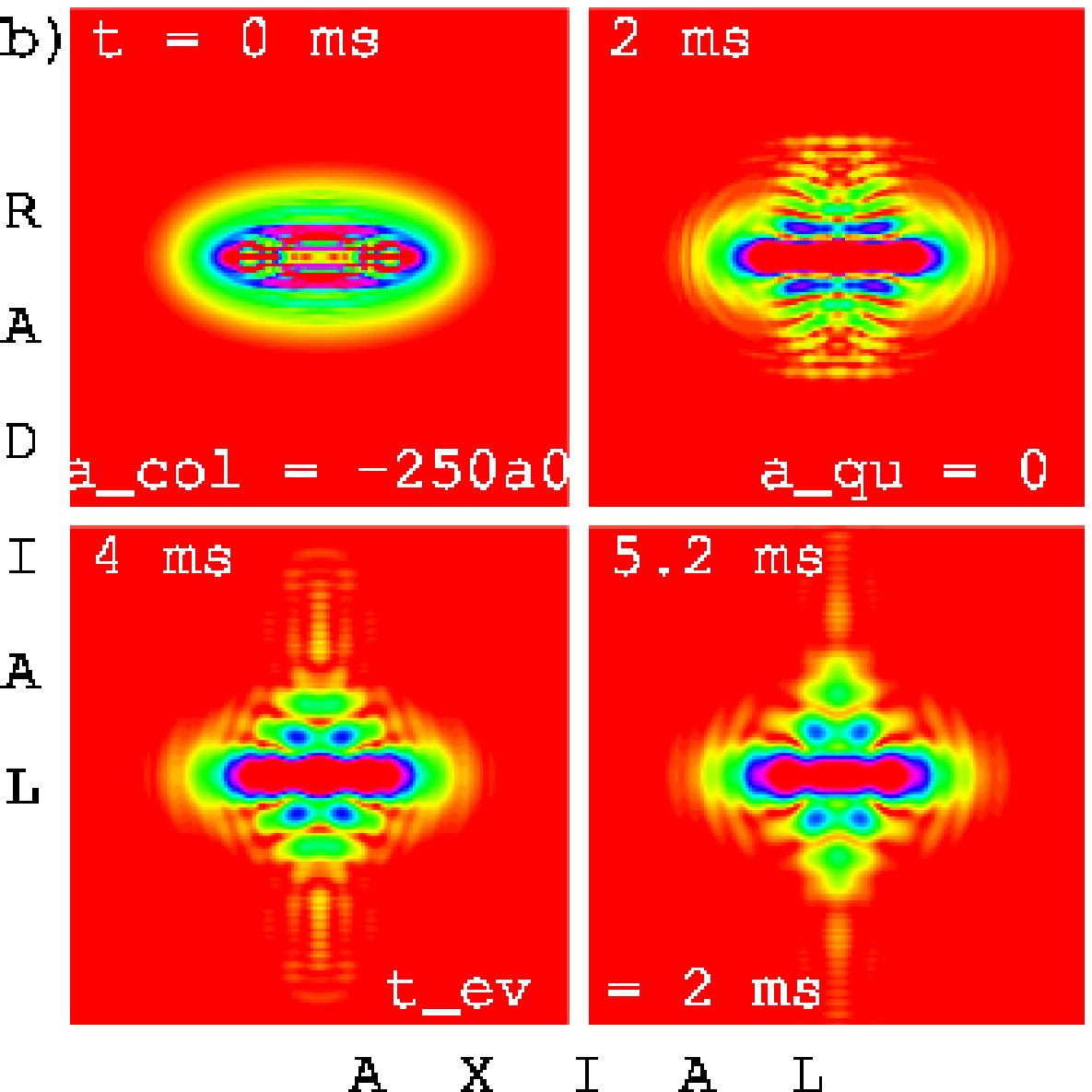}
\end{center}

\caption{A view of the evolution of radial jet 
on a mat of size 16 $\mu$m $\times$ 16 $\mu$m
 from a contour plot
of $|\phi(x,y,t)|^2$
for  
$a_{\mbox{initial}}=7a_0$,  $\xi=2$, 
$N_0=16000$, $a_{\mbox{quench}}= 0$,  and 
(a) $a_{\mbox{collapse}}=
-6.7a_0$, 
$t_{\mbox{evolve}} = 8$ ms, at times $t=0$, 4ms, 8ms
and
12 ms 
and (b) $a_{\mbox{collapse}}=
-250 a_0$, 
$t_{\mbox{evolve}} = 2$ ms, at times $t=0$, 2ms, 4ms
and
5.2 ms.}

\end{figure}

Donley {\it et al.} \cite{don} also considered expanding the condensate
before
observing the jet by jumping the scattering length to a large positive
value  $a_{\mbox{expand}} = 250 a_0$  after disrupting the
collapse. This procedure expands the size of the condensate so that it
might  be easier (or the only way)
to observe the jets in the laboratory. However, we find
that in all cases the jet is much less pronounced 
after this expansion. This is illustrated in figures 3 (a) and (b) where
we
plot the jet formation corresponding to  the cases reported in figures 1
(a) and (b), respectively, after expanding the condensate to
$a_{\mbox{expand}} = 250a_0$ as in the experimental result reported in 
figure 5 of \cite{don}. 
In  plots of figures 3 (a) and (b) the
condensate is of larger size than in the corresponding plots of figures 
1 (a) and (b). In figure 3 (a) the jet is almost destroyed. In figure 
3 (b) the jet  appears but it is wider due to expansion. 
We also expanded to $a_{\mbox{expand}} = 250a_0$
the jets for $a_{\mbox{collapse}} = -6.7a_0$ for different
$t_{\mbox{evolve}}$; the jets were almost 
destroyed in those cases.  After expanding to $a_{\mbox{expand}} = 250a_0$      
the jets for  $a_{\mbox{collapse}} = -250a_0$ became  wider due to
expansion but remained
visible.

In addition, we studied jet formation for  different values 
of $a_{\mbox{initial}}$ in place of $a_{\mbox{initial}}=7a_0$
and find that the scenario remains very similar independent of the initial
scattering length. However, the number of particles in the jet gives a
quantitative measure of jet formation and in the following 
we make a study of the number of atoms in the jet in different cases.

\begin{figure}
 
\begin{center}
\includegraphics[width=0.48\linewidth]{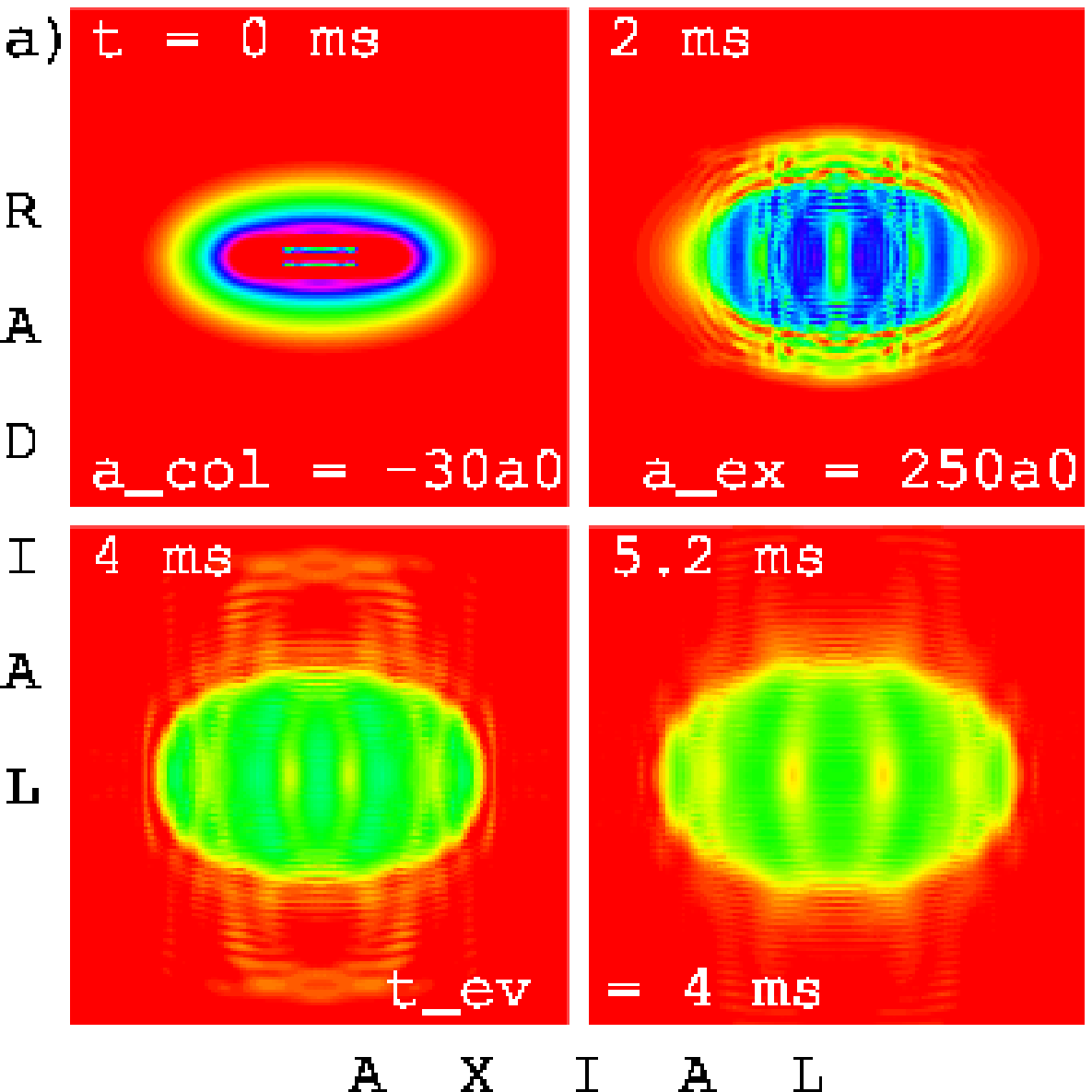}
\includegraphics[width=0.48\linewidth]{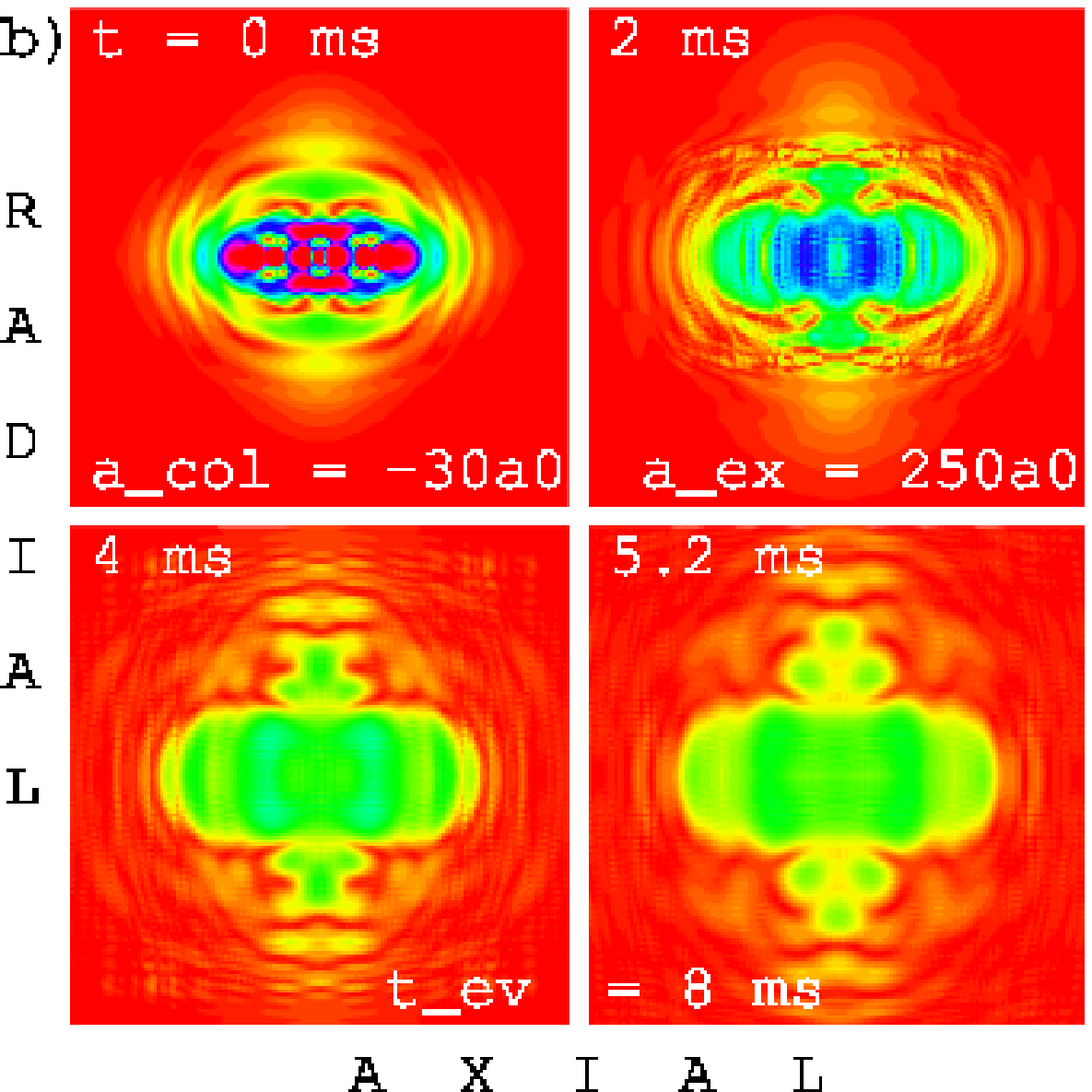}
\end{center}

\caption{A view of the evolution of radial jet at times $t=0$, 2ms, 4ms
and
5.2 ms on a mat of size 16 $\mu$m $\times$ 16 $\mu$m
from a contour plot
of $|\phi(x,y,t)|^2$
for  
$a_{\mbox{initial}}=7a_0$, $a_{\mbox{collapse}}=-30a_0$, $\xi=2$, 
$N_0=16000$, $a_{\mbox{expand}}= 250a_0$,  and (a) 
$t_{\mbox{evolve}} = 4$ ms and (b) $t_{\mbox{evolve}} = 8$ ms.}

\end{figure}

Next we calculated the number of jet atoms in different cases by
integrating the 
wave function over  the relevant region where the jet is formed. 
The normalization condition (\ref{5}) gives the total number of atoms 
in the condensate via $N_0 {\cal N}_{\mbox{norm}}$. After an examination of
figures 1  and 2 or other relevant jet figures
we separate the condensate at 5.2 ms (as in the
experiment) according to $x$ values into the central part and jet. The
$x$
integral in  (\ref{5}) is then separated into the central part and jet in
each case.   The jet part (outer $x$ values) of the integral
(\ref{5}) multiplied by $N_0$ gives the number of jet atoms. A similar
procedure has been used in \cite{th10} to calculate the number of jet
atoms. However, in actual experiment to see jet or any other phenomenon,
the
harmonic trap has to be removed and the condensate allowed to expand and
photographed. This  enlarges the condensate to be photographed
without presumably losing its actual shape and characteristics. Vortices
and dark and
bright solitons  photographed in this fashion give the true picture
of the condensate before free expansion. In numerical
simulation, on the other hand, it is possible and much easier to count the
jet atoms more accurately without any 
expansion. Assuming that there is not much experimental error in
counting the jet atoms after free expansion we attempt to compare the two
in the following.

For a fixed $N_0=16000$ and 
 $a_{\mbox{initial}}=7a_0$ 
we calculate the number of atoms in the jet 
for different
evolution time $t_{\mbox{evolve}}$ and $a_{\mbox{collapse}}$.
In none of the cases an expansion to $a_{\mbox{expand}}$ was applied.
The results are plotted in  figures  4. 
In figure 4 (a) we plot the variation of the number in jet vs. 
$t_{\mbox{evolve}}$
for  $a_{\mbox{collapse}}=-30a_0$ and in figure 4 (b)  we plot the variation of
the number in jet vs.       $|a_{\mbox{collapse}}|/a_0$  for 
 $t_{\mbox{evolve}}=4 $  ms. In figure 4 (a) we also plot the
experimental result for the  number of atoms in jet for the same values of the
parameters and find the agreement of our calculation with experiment to be
quite satisfactory. It is noted that  an expansion to $a_{\mbox{expand}}$
was not applied in the data reported in figure 6 (a) of \cite{don,don2}.

 In addition, in
figure 4 (a) we
compare the present results with the theoretical calculations by Calzetta
{\it et al.} \cite{th9} and Bao {\it et al.} \cite{th10}. The calculation
by Bao {\it et al.} \cite{th10} is essentially based on  mean-field 
GP equations as in this study and they correctly identify the jet atoms as
being a part of the condensate. They employ a fully asymmetric
mean-field model    in their description of jet formation. 
Yet the present  result for the number of atoms
in the jet is  larger  than theirs and in better agreement with
experiment \cite{don}. The reason for this discrepancy is
unknown.

The calculation by Calzetta {\it et al.} \cite{th9} uses a theoretical
model beyond mean-field  taking into consideration quantum
fluctuations. The number of jet atoms of Calzetta {\it et al.} \cite{th9}
is in good agreement with the experiment \cite{don} as well as with the
present study.  In figure 4 (a) we plot the average over the number of jet
atoms obtained by Calzetta {\it et al.} As the physical inputs and the
dynamics of the study of Calzetta {\it et al.} and the present study are
quite distinct, it is difficult to compare the two and  conclude about the
effect of quantum fluctuations on jet formation.  The effect of quantum
fluctuations could turn out to be significant in various aspects of the 
experiment of the collapse of a BEC including the formation of jet atoms. 
For example, they are of utmost relevance in the molecule formation in a
collapsing BEC near a Feshbach resonance and in subsequent
rapid atom-molecule
oscillation   as
observed recently \cite{don1}. Also, the burst atoms cannot be properly
described by a mean-field model and quantum corrections
could be significant \cite{th9,th10}.
Further studies are needed to identify
clearly the
effect of  quantum fluctuations on jet formation.

\begin{figure}
 
\begin{center}
\includegraphics[width=0.48\linewidth]{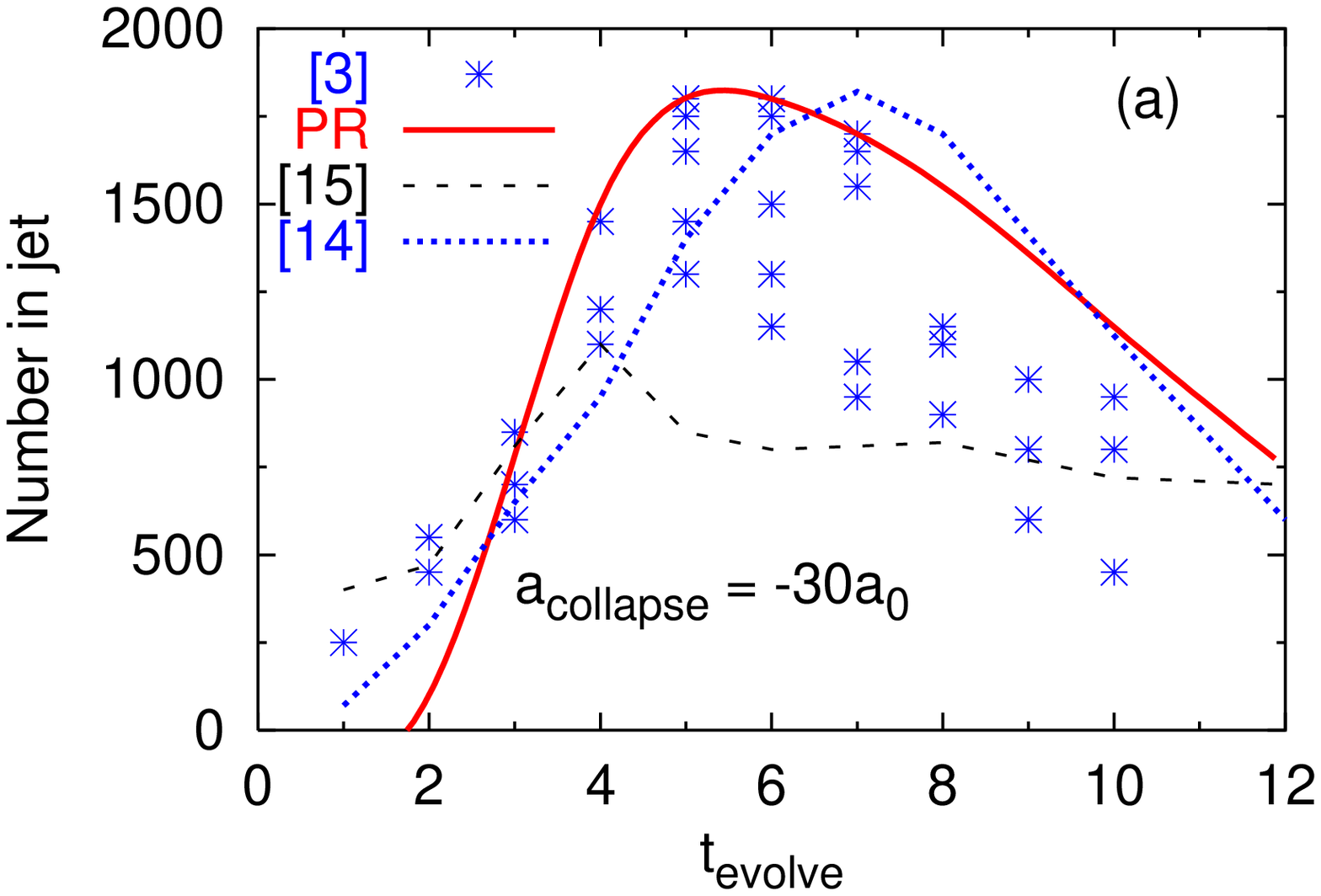}
\includegraphics[width=0.48\linewidth]{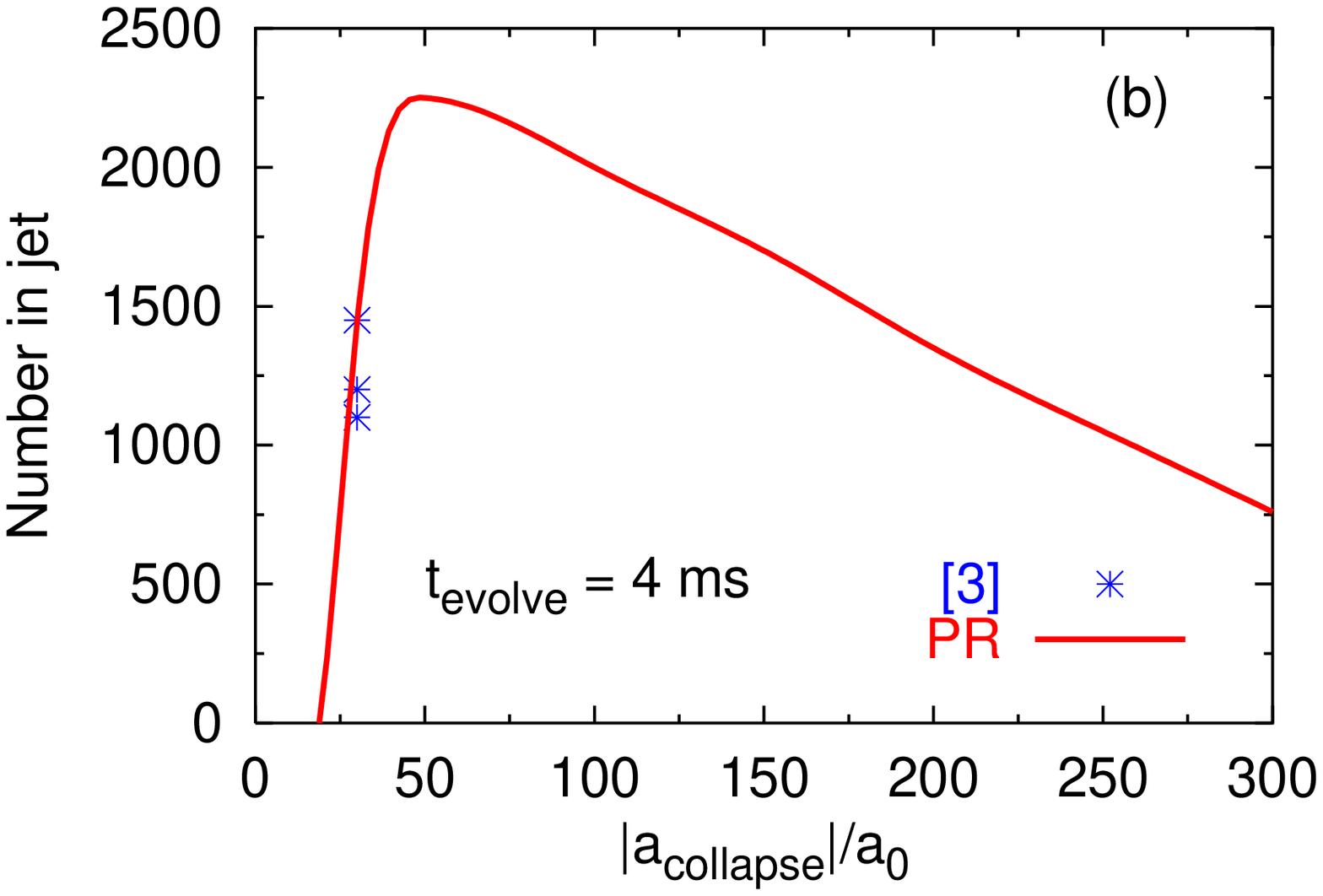}
\end{center}

\caption{Number of atoms in jet 5.2 ms after jumping the scattering length to 
 $a_{\mbox{quench}} = 0$
for
$N_0=16000$, $a_{\mbox{initial}}=7a_0$,
$\xi=2$  vs. (a)  $t_{\mbox{evolve}}$ for  $a_{\mbox{collapse}}=-30a_0$ and 
(b)  $|a_{\mbox{collapse}}|/a_0$  for  $t_{\mbox{evolve}}= 4$ ms; blue
star $-$ 
experiment  of Donley {\it et al.} \cite{don}, black dashed  line $-$ mean field 
model of  Bao  {\it et al.} \cite{th10}, blue dotted line $-$ average over
theoretical
results of Calzetta  {\it et al.} \cite{th9}, red full line
$-$ present  result.}

\end{figure}

The investigation by Saito {\it et al.} \cite{th3}  is very similar to
this
study in applying an axially-symmetric mean-field model and producing
the essentials of
jet formation. They also provide a physical     explanation of jet
formation. In
the collapsing condensate two distinct spikes are formed in the
condensate wave function along the axial
direction as the atomic interaction is changed from repulsive to
attractive.  These two spikes act as sources of matter waves and jet
is the interference pattern of matter waves from these two
sources \cite{th3}. 
However, we feel that 
small local
spike(s) in the wave function expand in the radial direction in the form
of
a jet
when the collapse is  stopped by
removing the atomic attraction and the force on the spike(s) is suddenly
changed
\cite{don}.

\section{Discussion}

In this section we give a physical explanation of the different types of
jet formation noted in last section.
The jet formation is more dramatic when the vigorous collapse of a
condensate is suddenly stopped by turning the attractive condensate
noninteracting ($a_{\mbox{quench}}=0$) or repulsive
($a_{\mbox{expand}}=250a_0$). In the strongly collapsing condensate 
local radial  spikes are formed during particle loss as can be
seen from a plot of the numerically calculated wave function
\cite{th2} and in experiment \cite{don}. 
During particle loss the top of the spikes 
are torn and ejected out and new spikes are formed until the
explosion and particle loss are over. There is a
balance between central atomic attractive force and the outward kinetic
pressure. If the attractive force is now suddenly removed by stopping the
collapse by applying $a_{\mbox{quench}}=0$, the highly collapsed 
condensate expands due to 
kinetic pressure,
becomes larger  and the recombination of atoms is greatly reduced. 
Consequently, the spikes expand and develop into a prominent jet
\cite{don}
for  $a_{\mbox{quench}}=0$ as in figure 1 (a). 

However, if the condensate
is expanded further
by applying $a_{\mbox{expand}}=250a_0$, in the cases studied, the spike as
well as the condensate expand so much that the prominent jet becomes in
general 
more diffuse as in figure 3 (b)
or completely destroyed as in figure 3 (a).  If the attractive condensate
is allowed to collapse for sufficiently long time, the explosion stops
eventually and a relatively cold remnant condensate is formed. At that
stage there would be almost no prominent  spikes in the wave function and
no jet could be
formed by applying   $a_{\mbox{quench}}=0$. With the increase of evolution
time of a collapsing condensate the jet becomes less prominent as
can be seen in figures 1 (a), (b) and 4 (a) and eventually disappears for
large 
$t_{\mbox{evolve}}$. However, the collapse and decay of particles start
after a finite  $t_{\mbox{evolve}}$ before which the jet formation is
practically absent as in figure 4 (a).  
If the condensate is weakly attractive as in figure 
2 (a) the collapse is also weak and the spikes are less
pronounced. Consequently, upon stopping the collapse a wide jet is formed
after a longer interval of time as in figure 2 (a).
As $|a_{\mbox{collapse}}|$ is increased at a fixed $t_{\mbox{evolve}} = 4$
ms as in figure 4 (b) one gradually passes from a strongly collapsing
condensate to a relatively cold remnant, that is from a region of
prominent jet formation to a region with less prominent jet. For smaller 
$|a_{\mbox{collapse}}|$ the collapse is weaker and jet formation is absent
as in figure 4 (b).

In the actual experiment \cite{don} the jet is found not to possess axial
symmetry. In the present study we use an axially symmetric model to
describe the essential features of the jet.  Hence, 
although  an axially symmetric
model is enough for a qualitative description of the jet, a full
three-dimensional model might be necessary for its complete quantitative
description.

\section{Conclusion}

In conclusion, we have employed a numerical simulation based on the
accurate
solution \cite{sk1}
of the mean-field Gross-Pitaevskii equation with a cylindrical
trap to study the jet formation as observed in the
recent experiment of an attractive collapsing condensate by Donley {\it
et al.} \cite{don}. In the GP equation we include a
quintic
three-body nonlinear recombination loss term \cite{th1}
that accounts for the decay
of the strongly attractive condensate.  The result of the present
simulation is in good agreement with the experimental result for 
jet formation \cite{don}.
We also compare the present result with two other recent theoretical
calculations of jet formation \cite{th9,th10}. 
Of the different aspects of the experiment by Donley {\it et al.} the
dynamics 
of relatively hot emitted burst and missing atoms seems to be beyond
mean-field treatment \cite{th9,th10}. However, the various properties
of the cold
residual
condensate including jet formation seem to be describable by mean-field
models. In fact, many features of the experiment by 
by Donley {\it et al.}
\cite{don}, specially the detailed behavior of the surviving remnant
condensate \cite{th2,th3,th10}  and jet formation, 
have been  understood by introducing  the rather conventional
three-body recombination loss in the standard mean-field GP equation,
with a loss rate compatible with other
studies \cite{th2,th5,th10,k3,esry}.
 
 
\ack 

We thank Dr E. A. Donley for a helpful correspondence \cite{don2}.
The work is supported in part by the CNPq 
of Brazil.

\end{document}